\documentclass[%
 reprint,
superscriptaddress,
 amsmath,amssymb,
 aps,
]{revtex4-2}

\usepackage{graphicx}
\usepackage{dcolumn}
\usepackage{bm}
\usepackage{siunitx}
\usepackage{placeins}
\usepackage{hyperref}
\usepackage[capitalise]{cleveref}
\usepackage{braket}

\newcounter{milestonecount}

\newcommand{\VH}{V_\textrm{H}}
\newcommand{\VRFR}{V_\mathrm{RF}^\mathrm{R}}
\newcommand{\VRFL}{V_\mathrm{RF}^\mathrm{L}}
\newcommand{\VQD}{V_\mathrm{QD}}
\newcommand{\VCS}{V_\mathrm{CS}}
\newcommand{\dVCS}{\Delta V_\mathrm{CS}}
\newcommand{\VR}{V^\mathrm{R}_\mathrm{bias}}

\newcommand{\phiL}{\phi^\mathrm{L}_\mathrm{RF}}

\newcommand{\VLO}{V_\mathrm{LO}}
\newcommand{\VTL}{V_\mathrm{TL}}
\newcommand{\VTR}{V_\mathrm{TR}}
\newcommand{\VRO}{V_\mathrm{RO}}

\newcommand{\VHv}{V_\mathrm{H}}

\newcommand{\VCSv}{V_\mathrm{CS}}
\newcommand{\VTRv}{V_\mathrm{TR}}

\makeatletter
\def\@affilnum#1{\textsuperscript{\normalfont#1}}
\makeatother

\begin{document}

\title{Charge sensing the parity of an Andreev molecule}
\author{David~van~Driel}
\author{Bart~Roovers}
\author{Francesco~Zatelli}
\author{Alberto~Bordin}
\author{Guanzhong~Wang}
\author{Nick~van~Loo}
\author{Jan~Cornelis~Wolff}
\author{Grzegorz~P.~Mazur}
\affiliation{QuTech and Kavli Institute of NanoScience, Delft University of Technology, 2600 GA Delft, The Netherlands}
\author{Sasa~Gazibegovic}
\author{Ghada~Badawy}
\author{Erik~P.~A.~M.~Bakkers}
\affiliation{Department of Applied Physics, Eindhoven University of Technology, 5600 MB Eindhoven, The Netherlands}
\author{Leo~P.~Kouwenhoven}\email{l.p.kouwenhoven@tudelft.nl}
\author{Tom~Dvir}
\affiliation{QuTech and Kavli Institute of NanoScience, Delft University of Technology, 2600 GA Delft, The Netherlands}

\date{\today}

\begin{abstract}
The proximity effect of superconductivity on confined states in semiconductors gives rise to various bound states such as Andreev bound states (ABSs), Andreev molecules and Majorana zero modes.
While such bound states do not conserve charge, their Fermion parity is a good quantum number. One way to measure parity is to convert it to charge first, which is then sensed. In this work, we sense the charge of ABSs and Andreev molecules in an InSb-Al hybrid nanowire using an integrated quantum dot operated as a charge sensor.
We show how charge sensing measurements can resolve the even and odd states of an Andreev molecule, without affecting the parity.
Such an approach can be further utilized for parity measurements of Majorana zero modes in Kitaev chains based on quantum dots.
\end{abstract}

\maketitle

\section*{Introduction}

Majorana zero modes (MZMs) are predicted to appear at the ends of a 1D chain of spin-polarized, electronic sites with superconducting pairing~\cite{kitaev_unpaired_2001, oreg_helical_2010,lutchyn_majorana_2010}, which can be implemented using quantum dots (QDs) coupled to superconductors~\cite{sau_realizing_2012, fulga_adaptive_2013}. Even a minimal, two-site Kitaev chain hosts MZMs in a parameter sweet spot~\cite{leijnse_parity_2012, tsintzis_creating_2022}, and was recently realized in a semiconductor-superconductor hybrid nanowire~\cite{dvir_realization_2023}. 
MZMs in Kitaev chains are predicted to be robust to local perturbations and obey non-Abelian statistics, allowing for the demonstration of Ising anyon fusion rules and braiding~\cite{boross_braiding-based_2023, liu_fusion_2022}. Qubit states can be encoded in the parity of pairs of MZMs, making parity readout crucial for any quantum information experiment involving MZMs. Proposed readout techniques include circuit quantum electrodynamics~\cite{contamin_hybrid_2021, hinderling_flip-chip-based_2023}, quantum capacitance~\cite{plugge_majorana_2017} and charge sensing~\cite{aasen_milestones_2016}. To read out parity using a charge measurement, it must first be converted into charge, and then sensed~\cite{szechenyi_parity--charge_2020}. Although charge sensing has been applied to semiconductor-superconductor hybrids before, it has never been used to detect fractional charge differences~\cite{sabonis_comparing_2021, razmadze_radio-frequency_2019}. This can potentially be necessary for parity readout in Kitaev chains, given that the charge difference between the even and odd ground states can range from 0 to $1e$. In this work, we present charge sensing measurements of a hybrid semiconductor-superconductor system.
First, we measure the charge of an Andreev bound state (ABS) in its even and odd Fermion parity ground states. Then, we couple a QD to an ABS to form an Andreev molecule, and infer its ground state parity from tunnel spectroscopy and charge sensing measurements. As opposed to transport, charge sensing does not alter the parity itself, highlighting its potential as a tool for MZM parity readout in Kitaev chains.

\section*{Results}

\subsection*{Device characterization}

\begin{figure*}[ht!]
    \centering
    \includegraphics[width=\linewidth]{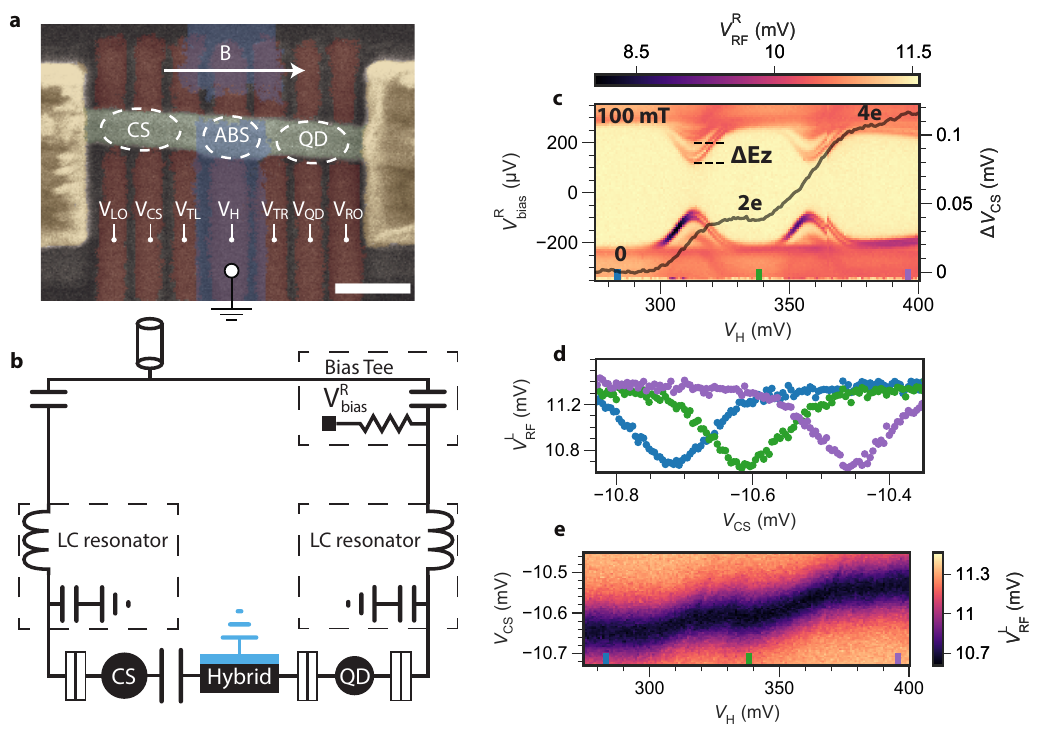}
    \caption{\textbf{Device set-up and characterization.} 
    \textbf{a.} False color SEM micrograph of our device. An InSb nanowire (green) is placed on an array of bottom gates (brown) and contacted by normal Cr/Au leads (yellow). We define a quantum dot (QD) to the left of the hybrid segment, and operate it as a charge sensor (CS). Another QD can be formed using the gates to the right of the hybrid segment. The names of the gate voltages are indicated in the respective gates. The scale bar is \SI{200}{\nm}.
    \textbf{b.} DC equivalent circuit of our device. Both normal leads of the hybrid segment are connected to off-chip LC resonators, which are multiplexed. Each LC resonator has a bias tee that can be used to bias the leads with respect to the grounded Al.
    \textbf{c.} Amplitude of the reflected RF signal of the right lead, $\VRFR$, for varying voltage on the hybrid plunger gate, $\VHv$ and the right bias, $\VR$, for an external magnetic field $B = \SI{100}{mT}$, $\VTR=0$ and $\VRO = \SI{500}{mV}$. \textbf{Superimposed line:} Shift of the gate voltage corresponding to the Coulomb resonance of the CS, $\dVCS$, for each $\VHv$ setting, extracted from e. The Zeeman splitting of the first ABS is indicated by $\Delta \mathrm{E}_\mathrm{Z}$. 0, $2e$ and $4e$ indicate additional charge accumulated on the hybrid.
    \textbf{d.} Amplitude of the reflected RF signal of the left lead, $\VRFL$, for varying voltage on the CS plunger gate, $\VCSv$, at fixed values of $\VH$ indicated by the colored bars in panels c and e.
    \textbf{e.} $\VRFL$ for varying $\VCS$ and $\VH$.
    }\label{fig: 1}
\end{figure*}

\cref{fig: 1}a shows an SEM image of the device measured in this work.
An InSb nanowire is placed on a thin layer of gate dielectric, below which are finger gates.
See Refs.~\cite{heedt_shadow-wall_2021, van_driel_spin-filtered_2022} for details on device fabrication.
The middle section of the wire, the hybrid segment, is contacted by a grounded Al thin film and hosts ABSs.
We define a QD to the left of the hybrid segment by setting $\VLO$ and $\VTL$ to create tunnel junctions in the nanowire. We control the QD's electrochemical potential using $\VCS$ and operate it as a single lead QD charge sensor (CS)~\cite{house_high-sensitivity_2016}.
The tunnel gate between the CS and hybrid segment is kept at a negative voltage $\VTL = \SI{-100}{mV}$ to fully quench transport and ensure their coupling is only capacitive.
The nanowire section to the right of the hybrid segment can be either a tunnel barrier or a QD, depending on the tunnel gate voltages $\VRO$ and $\VTR$. 

The nanowire is contacted by two normal Cr/Au leads that can be used for DC transport and radio frequency (RF) reflectometry measurements. \cref{fig: 1}b shows the DC equivalent circuit of the device and its connections. Both normal leads are connected to LC resonators with bias tees, allowing us to independently voltage bias them with respect to the grounded Al. The LC resonators are off-chip and multiplexed, see Ref.~\cite{hornibrook_frequency_2014} for details. Further details on the reflectometry circuit can be found in Ref.~\cite{wang_parametric_2022}; for resonator characterization, see~Sec.~\ref{sec:Resonator characterization} of the Supplemental Material~\cite{supplemental}. Experiments are performed in a dilution refrigerator at a base temperature of \SI{30}{\milli\kelvin}.

\subsection*{ABS even ground state charge}

The hybrid segment hosts ABSs that can have an odd, doublet ground state ($\{ \ket{\downarrow}, \ket{\uparrow} \}$) or an even ground state. We consider an ABS in the atomic limit, where the even ground state is a singlet: $\ket{S} =  u\ket{0}-v\ket{2}$, with 0 and 2 denoting the occupation of a single orbital~\cite{assouline_shiba_2017, baranski_-gap_2013}.
The singlet state changes between being mostly unoccupied (0-like) or doubly occupied (2-like) when the electrochemical potential of the ABS changes. Note that this is a gradual change of ground state, as opposed to changing a QD ground state by occupying additional electrons. When the semiconductor-superconductor coupling is stronger than the charging energy, the ABS has an even ground state with a doublet excited state~\cite{lee_spin-resolved_2014}.


 
To characterize the hybrid segment, we set $\VRO = \SI{500}{mV}$ to accumulate electrons, and create a tunnel barrier with $\VTR$ to perform tunneling spectroscopy from the right lead.
\cref{fig: 1}c shows amplitude of the reflected RF signal of the right lead, $\VRFR$, for varying hybrid plunger gate voltage, $\VHv$, and right voltage bias, $\VR$. A dip in $\VRFR$ can be related to a peak in differential conductance~\cite{wang_parametric_2022}.
The hybrid segment has a hard-gapped density of states with two well-separated ABSs of which the energy can be controlled with $\VH$.
Each ABS excitation is split into two resonances, because we apply $\SI{100}{mT}$ along the nanowire axis.
From the Zeeman splitting $\Delta E_\mathrm{Z} = \SI{80}{\micro\electronvolt}$, we obtain an effective $g$-factor of $g = 13.8$ for the ABS.
The ABS excitations do not cross zero bias, signalling an even ground state for the entire $\VH$ range.

\cref{fig: 1}d shows amplitude of the reflected RF signal of the left lead, $\VRFL$, for varying plunger gate voltage of the CS, $\VCSv$, taken at three different values of $\VHv$.
A Coulomb resonance of the CS is seen as a dip in $\VRFL$ whenever a CS level is aligned with the Fermi level of the left lead~\cite{persson_excess_2010}.
We see that the $\VCS$ value corresponding to the $\VRFL$ minimum changes due to $\VH$, which has two possible causes. 
First, $\VH$ directly gates the CS due to cross-capacitance. Second, $\VH$ can change the occupation of the hybrid segment and the resulting charge is sensed by the CS. 
To extract the charge sensing signal only, we define virtual gates that are linear combinations of the physical gate voltages, to compensate for cross-capacitances between gates (for details see Sec.~\ref{sec:virtual gates} of the Supplemental Material~\cite{supplemental}).

In \cref{fig: 1}e, we show one particular CS Coulomb resonance for varying $\VCSv$ and $\VHv$. We have subtracted a global slope from the data, which we attribute to imperfect virtual gate settings. The global slope was chosen such that the CS resonance is roughly constant in $\VH$ up to $\SI{300}{\milli\volt}$, for which there are no sub-gap states (for details and raw data, see Sec.~\ref{sec:virtual gates} of the Supplemental Material~\cite{supplemental}.

The superimposed black line of \cref{fig: 1}c shows the shift of the CS Coulomb resonance, $\dVCS$, as found from \cref{fig: 1}e.
We see that $\dVCS$ depends on $\VH$ and changes most strongly when there is an ABS at sub-gap energies. The absence of sharp jumps in $\dVCS$ shows that there are no abrupt changes of charge, which is consistent with an even ground state for the entire $\VH$ range as found from \cref{fig: 1}c. We interpret the change in $\dVCS$ at $\VH = \SI{310}{\milli\volt}$ and $\VH = \SI{360}{\milli\volt}$ as the CS sensing the charge of the ABSs changing continuously from 0 to $2e$, where $e$ is the charge of the electron.

Gradual change of charge without change of parity, as seen here, has been observed before in normal double quantum dots~\cite{dicarlo_differential_2004, hu_gesi_2007}. In our case, however, the ABS exchanges charge with a large Al reservoir and becomes a coherent superposition of 0 and $2e$.





\subsection*{Single ABS parity readout}
\begin{figure}
    \centering
    \includegraphics[width=\columnwidth]{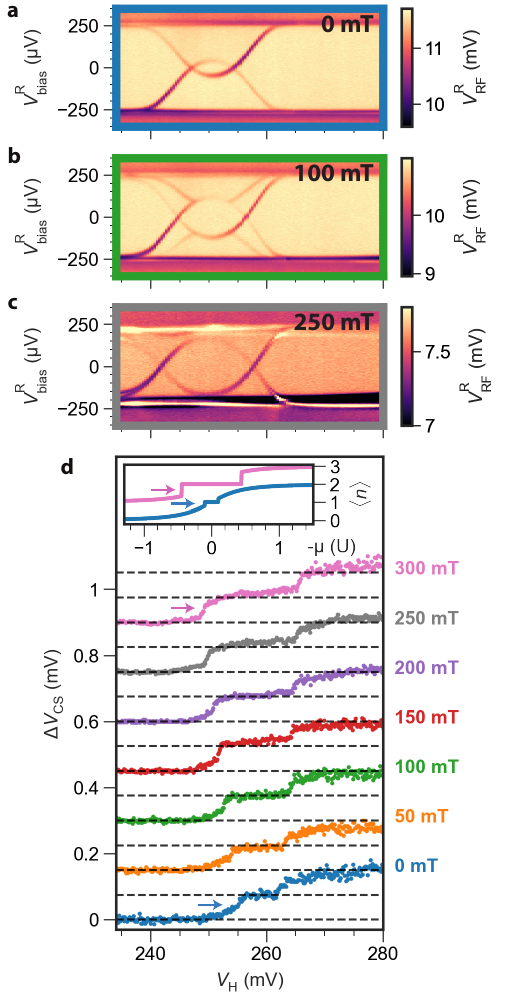}
    \caption{\textbf{Charge sensing the even and odd ground states of an ABS.} 
    \textbf{a.-c.} Amplitude of the reflected RF signal of the right lead, $\VRFR$, for varying hybrid gate voltage $\VHv$ and right bias $\VR$, taken at $B = \SI{0}{mT}$ (a.), $B = \SI{100}{mT}$ (b.) and $B = \SI{250}{mT}$ (c.) for $\VTRv = \SI{0}{mV}$, $\VTL=\SI{-100}{mV}$, $\VLO=\SI{170}{mV}$ and $\VRO = \SI{500}{mV}$.
    \textbf{d.} Shift of the CS Coulomb resonance, $\dVCS$, at different $B$ values for $\VTL=\SI{-140}{mV}$ and $\VLO=\SI{172}{mV}$. The horizontal lines are offset by $\SI{75}{\micro\V}$, which corresponds to a charge difference of approximately $1e$. The arrows indicate the even-odd transition at $B = \SI{0}{mT}$ and $B = \SI{300}{mT}$ respectively.
    \textbf{inset.} The occupation expectation value of the ABS ground state, $\langle n \rangle$, for varying chemical potential, $\mu$, at different $B$ values, calculated in the atomic limit. The arrows indicate the even-odd transition.
    }
    \label{fig: 2}
\end{figure}

To use our charge sensor for parity readout, we focus on an ABS with a ground state that can be changed from even to odd parity using $\VH$. \cref{fig: 2}a shows spectroscopy of such an ABS in a different $\VH$ range than \cref{fig: 1}, for $B = \SI{0}{mT}$. We see a sub-gap state that crosses zero energy twice for changing $\VH$, give rising to a characteristic ABS eye-shape~\cite{deng2016majorana, valentini2021nontopological}. The ABS ground state between the zero-energy crossings is a spin doublet $\ket{D}=\ket{\downarrow}, \ket{\uparrow}$~\cite{lee_spin-resolved_2014}. Having a doublet ground state at $B = \SI{0}{mT}$ signals that the ABS has a charging energy that is non-negligible. While this is uncommon among most ABSs in the hybrid segment, we utilize it to measure varying parities (for measurements in a larger $\VH$ range, see Sec.~\ref{sec:spectrum} of the Supplemental Material~\cite{supplemental}). \cref{fig: 2}b and c show spectroscopy of the ABS at $B = \SI{100}{mT}$ and $B = \SI{250}{mT}$, respectively. The magnetic field Zeeman splits the doublet states, making the odd state lower in energy than the even state over a larger $\VH$ range.

In \cref{fig: 2}d we show the processed shift of the CS Coulomb resonance, $\dVCS$, for varying $\VH$ at different $B$ values. To obtain $\dVCS$ for each $B$ value, we first set the CS on Coulomb resonance using $\VCS$ and measure $\VRFL$ for varying $\VH$. We then convert the measured $\VRFL$ to the corresponding shift of the CS Coulomb resonance, resulting in $\dVCS$. For details, see Sec.~\ref{sec:fig2 data processing} of the Supplemental Material~\cite{supplemental}.

For each $B$ value, we identify three distinct $\VHv$ regions where the $\dVCS$ response is roughly flat. The middle region of these around $\VHv \approx \SI{260}{mV}$ occupies a larger $\VH$ range for increasing $B$. This is consistent with $\ket{\downarrow}$ being the ground state for a larger $\VH$ range, as inferred from \cref{fig: 2}a-c (note that the $\VH$ values for which the ABS has an odd ground state differ slightly between panels a-c.\ and d.\ due to different $\VLO$ and $\VTL$ gate voltages, which were chosen to optimize the sensitivity of the charge sensor). We interpret the three distinct $\dVCS$ values for increasing $\VH$ as corresponding to the 0-like even, singly occupied odd, and 2-like even states of the ABS.






We observe finite curvature in $\dVCS$ for the even ground state, which we attribute to mixing of the 0 and 2 occupations, similar to \cref{fig: 1}e. The curvature is most visible close to the even-odd transition, and becomes less apparent with increasing $B$. At $B = \SI{0}{mT}$, the even-odd transition (indicated by the blue arrow) occurs when the ABS is near its energy minimum, where the average ABS charge is $1e$ for both the even and odd states~\cite{danon_nonlocal_2020, menard_conductance-matrix_2020}. The even state charge gradually changes from 0 to almost $1e$ in the $\VH$ range before the transition.
At $B = \SI{300}{mT}$, the even-odd transition occurs when the ABS is no longer close to its energy minimum. The difference in charge between the even and odd states is then greater, resulting in a sharper change of $\dVCS$, as indicated by the pink arrow. 

If the curvature of $\dVCS$ is exclusive to the even ground state, we can use it to infer the parity of the ABS using charge sensing only. To illustrate this point, we calculate and show the charge of an ABS in the inset of \cref{fig: 2}d. Here, the average occupation, $\langle n \rangle$, is shown for varying chemical potential, $\mu$, at two different values of $B$ (for more details see Sec.~\ref{sec:atomic limit} of the Supplemental Material~\cite{supplemental}). We conclude that at low $B$, curvature in $\langle n \rangle$ is a characteristic sign of an ABS in a singlet ground state, provided the parent gap is much larger than the charging energy.

So far, we have established that charge sensing measurements can resolve the charge differences between the even and odd ground states of a single ABS. In a two-site Kitaev chain, however, it is the combined parity of two hybridized QDs that has to be detected. We create a proxy system by hybridizing a QD with an ABS and charge sense it to give a minimal demonstration of parity readout for Kitaev chains. Parity readout of Majorana zero modes requires extra steps, which we detail in the discussion. We note that an alternating ABS-QD array can also constitute a Kitaev chain~\cite{Miles2023-dc, Samuelson2023-xh}.



\subsection*{Andreev molecule spectroscopy}

\begin{figure*}[ht!]
    \centering
    \includegraphics[width=\linewidth]{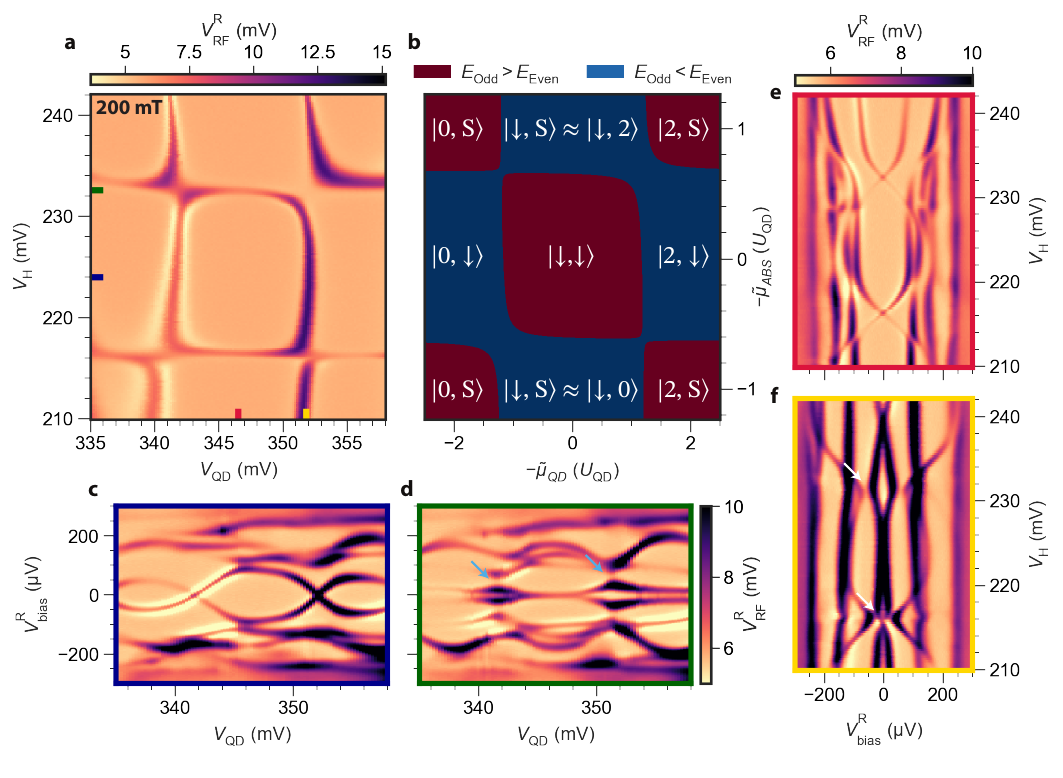}
    \caption{\textbf{Spectroscopy of a hybridized ABS-QD system.} 
    \textbf{a.} Amplitude of the reflected RF signal of the right lead, $\VRFR$ for varying QD plunger gate voltage, $\VQD$, and $\VH$ taken at $\VR=\SI{0}{V}$, $B = \SI{200}{mT}$, $\VTR = \SI{215}{mV}$ and $\VRO = \SI{340}{mV}$.
    \textbf{b.} 
    Parity of the lowest energy eigenstate of a coupled ABS-QD system for varying QD and ABS chemical potential $\tilde{\mu}_\mathrm{QD}$ and $\tilde{\mu}_\mathrm{ABS}$, computed in the atomic limit. We designate the ground state parity as even if $E_\mathrm{Odd}>E_\mathrm{Even}$ and odd if $E_\mathrm{Odd}<E_\mathrm{Even}$. The superimposed text indicates the ground states in the $\ket{N, M}=\ket{N}_\mathrm{QD} \otimes \ket{M}_\mathrm{ABS}$ basis of the ABS-QD system. We denote the ABS singlet state by $\ket{S}$, and highlight its dominant occupation for the $\ket{\downarrow, S}$ state. 
    \textbf{c., d.} $\VRFR$ for varying $\VQD$ and $\VR$, taken at two different values of $\VH$ indicated by the blue and green horizontal bars in panel a.
    \textbf{e., f.} $\VRFR$ for varying $\VH$ and $\VR$, taken at two different values of $\VQD$ indicated by the red and yellow vertical bars in panel a.
    }
    \label{fig: 3}
\end{figure*}

 To create the ABS-QD system, we define a QD to the right of the hybrid segment by creating a tunnel barrier using $\VRO$ (see \cref{fig: 1}a). We then lower the tunnel barrier between the QD and the hybrid segment by increasing $\VTR$.

\cref{fig: 3}a shows a charge stability diagram measured in $\VRFR$ for varying $\VH$ and QD plunger gate, $\VQD$. We see avoided crossings that indicate hybridization of the QD and ABS.
The QD resonance can be seen when the ABS is off-resonance (blue bar), which we attribute to local Andreev reflection on the QD and the usage of RF reflectometry instead of differential conductance (for details see Sec.~\ref{sec:DC fig 3} of the Supplemental Material~\cite{supplemental}).
Similarly, the ABS can be observed when the QD is off-resonance (e.g., red bar). Because of the strong hybridization between the QD and ABS, the latter can always be seen in spectroscopy upon probing the QD.

\cref{fig: 3}b shows the ground state parity of a QD coupled to an ABS, computed in the atomic limit. States are indicated in the ABS-QD number basis, with $\ket{S}$ denoting the ABS singlet (for details on the model, see Sec.~\ref{sec:atomic limit} of the Supplemental Material~\cite{supplemental}).  We emphasize that the singlet is dominantly 0-like or 2-like at $-\tilde{\mu}_\mathrm{ABS} = \pm 1\ U_\mathrm{QD}$. We see that \cref{fig: 3}b qualitatively reproduces most of the features of \cref{fig: 3}a.


 \cref{fig: 3}c and d show spectroscopy of the QD for varying $\VQD$ and $\VR$ at fixed values of $\VH$ indicated by the horizontal lines in \cref{fig: 3}a.
 When the ABS in the hybrid segment is not at zero-energy (blue bar, panel c), we see multiple sub-gap states, where the lowest one forms a typical eye-shape. This indicates the formation of Yu-Shiba-Rusinov (YSR) states on the QD, due to its strong coupling to the hybrid segment~\cite{yu_bound_1965, shiba_classical_1968, rusinov_superconductivity_1969, grove-rasmussen_superconductivity-enhanced_2009, jellinggaard_tuning_2016}. 
\cref{fig: 3}d (green bar) shows spectroscopy of the QD when the ABS excitation is at zero energy, i.e., on-resonance.
Here we see that the ABS and QD states form sub-gap bonding and anti-bonding states when they are on resonance, indicated by the blue arrows. Two hybridized sub-gap states are often referred to as an ``Andreev molecule'', which usually designates two coupled YSR states~\cite{steffensen_direct_2022, su_andreev_2017, kurtossy_andreev_2021, scherubl_transport_2019, estrada_saldana_two-impurity_2020, grove-rasmussen_yushibarusinov_2018, potts_large-bias_2023} or phase-tunable ABSs in Josephson junctions~\cite{haxell_demonstration_2023, coraiola2023hybridisation, kocsis_strong_2023}. Because our ABS-QD system shows hybridization of an ABS in a hybrid with a YSR state in a QD, we categorize it as an Andreev molecule. 

\cref{fig: 3}e and f show spectroscopy of the QD for varying $\VH$ at fixed values of $\VQD$ indicated by the vertical lines in \cref{fig: 3}a.
When the QD is off-resonance (red bar, panel e) the lowest sub-gap state crosses zero energy twice, and the excited states are close in energy to the superconducting gap.
\cref{fig: 3}f (yellow bar) shows spectroscopy when the QD excitation is at zero-energy.
When the ABS is on resonance, it splits the zero-bias peak of the QD, indicated by the white arrows. This effect is known to occur for QDs strongly coupled to ABSs~\cite{prada_measuring_2017, clarke_experimentally_2017, poschl_nonlocal_2022}.

\subsection*{Andreev molecule parity readout}

\begin{figure}[ht!]
    \centering
    \includegraphics[width=\columnwidth]{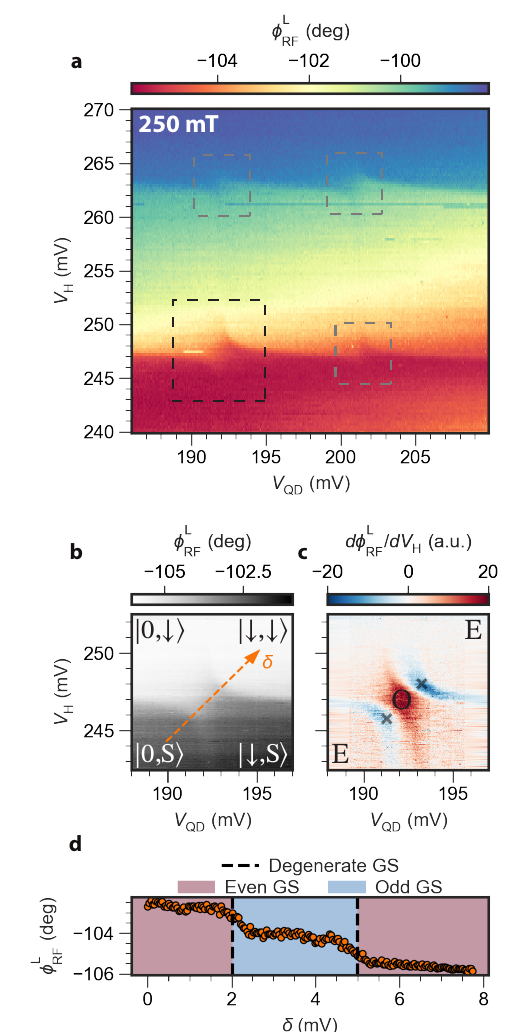}
    \caption{\textbf{Charge sensing measurements of the ABS-QD system.} All data was taken for $\VTR=\SI{255}{mV}$, $\VRO = \SI{230}{mV}$ and $B = \SI{250}{mT}$. 
    \textbf{a.} Phase of the reflected RF signal of the left lead, $\phiL$, for varying $\VQD$ and $\VH$. 
    \textbf{b.} Zoom in of a. for the range indicated by the dashed rectangle. The superimposed text indicates the ground state of the ABS-QD system in the same basis as in \cref{fig: 3}b.
    \textbf{c.} The numerical derivative, $d\phiL/d \VH$, of b. The superimposed text indicates the parity of the states in panel b, where "O" and "E" indicate an odd and even ground state, respectively. The two crosses indicate negative peaks of $d\phiL/d \VH$ found from peak-finding for the cut along the orange diagonal line shown in panel b.
    \textbf{d.} $\phiL$ for a cut along the orange diagonal line shown in panel b. Here, $\VQD$ and $\VH$ are changed in parallel, indicated by $\delta$. The vertical black lines correspond to the $\VQD$ and $\VH$ values of the black crosses in panel c, from which we infer a parity change. The red and blue shading indicate even and odd ground states, respectively.
    }\label{fig: 4}
\end{figure}

Next, we sense the charge of the ABS within the ABS-QD system. \cref{fig: 4}a shows the phase of the reflected RF signal of the left lead, $\phiL$, for varying $\VQD$ and $\VH$, at different gate and field settings from \cref{fig: 3}.
We set the CS on Coulomb resonance using $\VCS$, at the $\VQD$ and $\VH$ settings that correspond to the bottom left corner of \cref{fig: 4}a. Globally, we observe three horizontal regions where $\phiL$ depends almost only on $\VH$, which we interpret as the 0-like even, singly occupied $\ket{\downarrow}$ and 2-like even states of the ABS. In addition, there are 4 regions where $\VRFL$ depends visibly on $\VQD$, which we attribute to the QD hybridizing with the ABS. These regions are highlighted using dashed rectangles.

In \cref{fig: 4}b we show a zoom-in of panel a corresponding to the black rectangle. Here, the CS was gated to be on the steepest slope of the Coulomb peak using $\VCS$ at the start of the measurement. We see an off-diagonal avoided crossing, which signals an interdot transition~\cite{petta2005coherent}. Because this is the first interdot transition of this QD orbital and ABS, we label the bottom left corner of \cref{fig: 4}b with $\ket{0, S}$, where the QD orbital is unoccupied and the ABS is in the 0-like singlet state. The other states are labeled by counting the added electrons. Although the QD is proximitized by coupling to the hybrid, as seen in \cref{fig: 3}c, we retain the number basis for clarity.

\cref{fig: 4}c shows the numerical derivative of panel b, $d\phiL/d \VH$, after processing with a Savgol filter. The dips in $d\phiL/d \VH$ form two blue hyperbolas that indicate where the even and odd ground states are degenerate~\cite{hu_gesi_2007}. Based on the states inferred from panel b, we label the region between the hyperbolas as odd ("O") and the two areas outside as even ("E"). 

\cref{fig: 4}d shows a linecut of $\phiL$ taken along the orange line in panel b, where $\VQD$ and $\VH$ are changed in parallel, indicated by $\delta$. The two vertical, dashed black lines correspond to the gate values of the superimposed crosses in panel c, which we use to divide the range of $\delta$ into even and odd ground states. We note that there are three regions where $\phiL$ is roughly constant, and that these coincide with even and odd parity sectors. From this we conclude that we can read out the parity of the ABS-QD system by measuring its charge.

Extending parity readout using charge sensing to a two-site Kitaev chain requires an additional step however. In the "Poor Man's Majorana" sweet spot, the degenerate even and odd parity ground states have the same charge~\cite{tsintzis_creating_2022}. Detuning one QD creates a charge difference between the even and odd ground states on the other QD, while the states remain degenerate~\cite{tsintzis2023roadmap}. A local charge sensing measurement on the non-detuned QD can then tell the parity based on charge. This protocol of detuning and measuring charge has to be performed within the quasi-particle poisoning time, otherwise the parity flips during the measurement.  We note that our RF integration time of $\tau = \SI{9.3}{\milli\s}$ is on the order of quasiparticle poisoning times typically found in hybrid systems, although estimates vary strongly depending on device design~\cite{uilhoorn2021quasiparticle,bargerbos2022singlet}. This results in the CS predominantly sensing the average occupation of our system. We show SNR values for different integration times in Sec.~\ref{sec:SNR} of the Supplemental Material~\cite{supplemental}.





\section*{Conclusion}

In conclusion, we have operated a QD as a charge sensor and measured the charge of ABSs in a hybrid semiconductor-superconductor nanowire. We have found that the charge of an ABS can change by $2e$ while remaining in the even ground state. The charge difference between the even and odd states can be less than $1e$ due to the ABS exchanging charge with the superconductor. We have coupled a QD to an ABS to form a hybridized state, which we categorize as an Andreev molecule. Using the charge sensor, we can infer the parity of the ABS-QD system when the even and odd ground states have different charges. For a two-site Kitaev chain, both parities have the same charge in the sweet spot, and require an additional step for parity-to-charge conversion. We demonstrate that charge sensing can be used for parity readout in hybrid systems and is promising for usage in Kitaev chains. 

\FloatBarrier
\bibliography{Zotero}

\section*{Acknowledgements}
This work has been supported by the Dutch Organization for Scientific Research (NWO) and Microsoft Corporation Station Q. We wish to acknowledge useful discussions with Chun-Xiao Liu and Anasua Chatterjee.

\section*{Author contributions}
DVD, GW, AB, NvL, FZ, GM fabricated the devices. DVD and BR performed the electrical measurements. DVD, TD and BR designed the experiment and analyzed the data. DVD, TD and LPK prepared the manuscript with input from all authors. TD and LPK supervised the project. 

\section*{Data availability}
All raw data in the publication and the analysis code used to generate figures are available at \url{https://doi.org/10.5281/zenodo.10067038}.

\clearpage

\section*{Supplemental Material}
\renewcommand{\thesubsection}{\Roman{subsection}} 
\setcounter{figure}{0}
\renewcommand\thefigure{S\arabic{figure}}

\subsection{Resonator Characterization}\label{sec:Resonator characterization} 

\begin{figure*}[ht!]
    \centering
    \includegraphics[width=\linewidth]{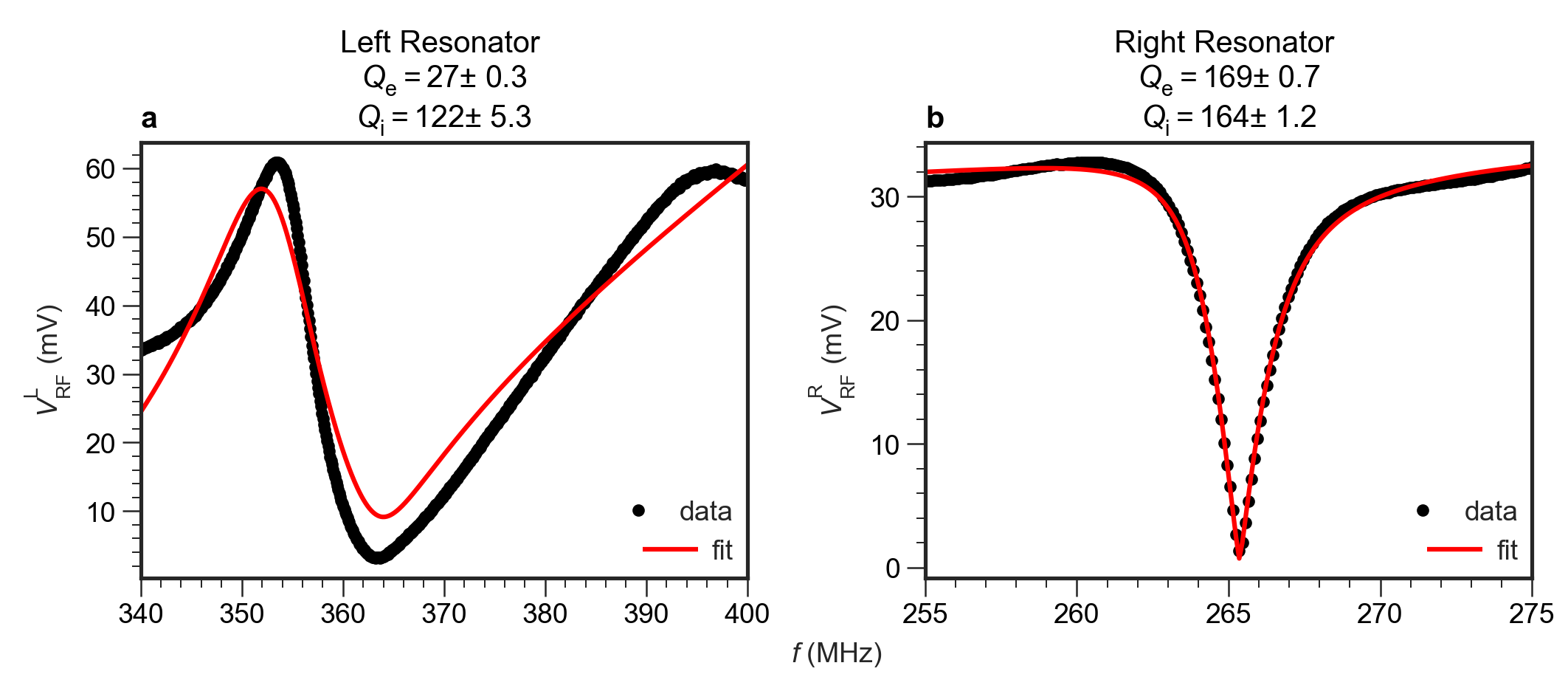}
    \caption{\textbf{Resonator characterization.} Fits of off-chip resonators performed according to the model described in Ref.~\cite{khalil_analysis_2012}.
    \textbf{a.} Amplitude of the reflected RF signal of the left lead, $\VRFL$, versus the frequency of the RF output signal, $f$. 
    \textbf{b.} Amplitude of the reflected RF signal of the right lead, $\VRFR$, versus the frequency of the RF output signal, $f$. 
    }\label{S: resonator}
\end{figure*}

Both normal contacts of the device presented in this work are bonded to off-chip LC resonators. \cref{S: resonator}a,b show the amplitude of the reflected RF signal of the left and right lead $V^\mathrm{L/R}_\mathrm{RF}$ for varying frequency, $f$, at $B = \SI{0}{mT}$. We fit the resonator responses using a model for asymmetric resonances by Khalil et al.~\cite{khalil_analysis_2012}. From the fit, we obtain the internal and external Q-factors listed in the figure. For all other measurements in this work, we fix the frequencies at $f_\mathrm{L} = \SI{364}{\MHz}$ and $f_\mathrm{R} = \SI{264}{\MHz}$. All RF measurements were performed using a Zurich Instruments UHFLI.

\subsection{Virtual Gates and \cref{fig: 1} Data Processing}\label{sec:virtual gates} 

\begin{figure*}[ht!]
    \centering
    \includegraphics[width=\columnwidth]{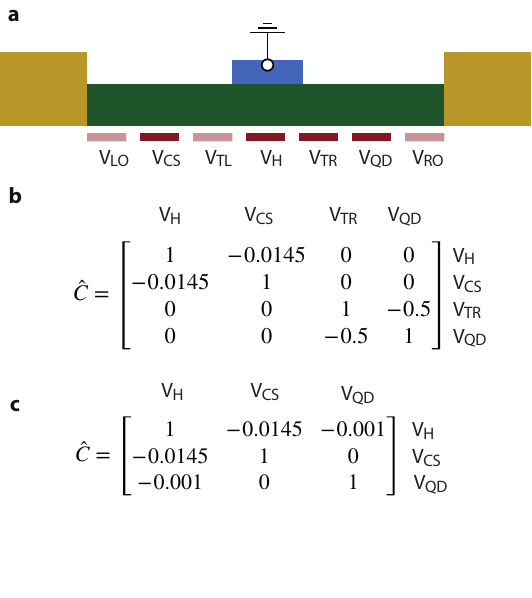}
    \caption{\textbf{Virtual gates.} 
    \textbf{a.} Device sketch that highlights the physical gates used to define the new virtual gates
    \textbf{b.} The cross-capacitance matrix for the gates highlighted in a. Off-diagonal numbers indicate a finite cross-coupling between physical gates, which is corrected for using the virtual gates. This setting is used for Fig.~1 and 2 .
    \textbf{c.} Same as panel b, but corresponding to the data from figure 4 of the main text.
    }\label{S: virtualgates}
\end{figure*}

\cref{S: virtualgates} shows the gates and cross-capacitance matrix $\hat{C}$ used for the virtual gates in this work. The matrix $\hat{C}$ allows us to estimate the actual electrochemical potential change due to cross-capacitance, using: $\vec{V}^\prime = \hat{C} \vec{V}$. Here, the vector $\vec{V}$ contains all the physical gate voltages and $\vec{V}^\prime$ is the resulting voltage on parts of the device. This allows us to change gates in a way that, for example, affects only the QD and not the CS electrochemical potential. For more details on virtual gates, see Refs.~\cite{volk_loading_2019, mills_shuttling_2019}.

\begin{figure*}[ht!]
    \centering
    \includegraphics[width=\linewidth]{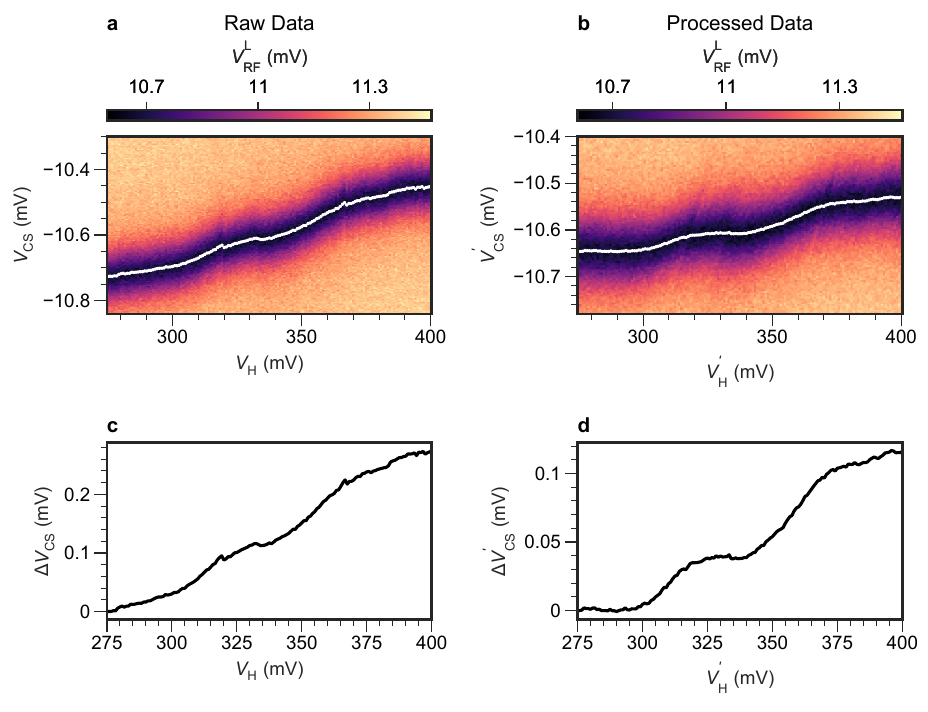}
    \caption{\textbf{Data processing of \cref{fig: 1}.} 
    \textbf{a.} Amplitude of the reflected RF signal of the left lead, $\VRFL$, for varying $\VCS$ and $\VH$. White markers indicate the minimum of the Coulomb resonance for each $\VH$ value.
    \textbf{b.} Same as a., but after removing a global slope between $\VCS$ and $\VH$. 
    \textbf{c.,d.} Shift of the CS Coulomb resonance, $\dVCS$, for each $\VHv$ setting, extracted from a. and b. respectively. These correspond to the white markers in panels a and b. The line of panel d is superimposed on \cref{fig: 1}c.
    }\label{S: data_process_fig1}
\end{figure*}

Our virtual gate implementation was imperfect, leading to a residual cross-capacitance between $\VH$ and $\VCS$. \cref{S: data_process_fig1}a shows that the CS resonance is also gated by $\VH$. We attribute the residual cross-coupling to the non-linearity of the device and an inaccurate choice of $\hat{C}$. We correct for this in post-processing by subtracting a global slope, until the CS resonance does not depend on $\VH$ up to $\VH = \SI{300}{\milli\V}$. This result is presented in \cref{fig: 1}e and in \cref{S: data_process_fig1}b. In \cref{S: data_process_fig1}d we show the minima of the Coulomb dip for each $\VH$ value, extracted from panels a., b. The processed data is seen to be devoid of a global slope. The line in \cref{S: virtualgates}d is superimposed in \cref{fig: 1}c.

\subsection{Extended Range Hybrid Spectrum and Charge Sensing}\label{sec:spectrum} 

\begin{figure*}[ht!]
    \centering
    \includegraphics[width=\linewidth]{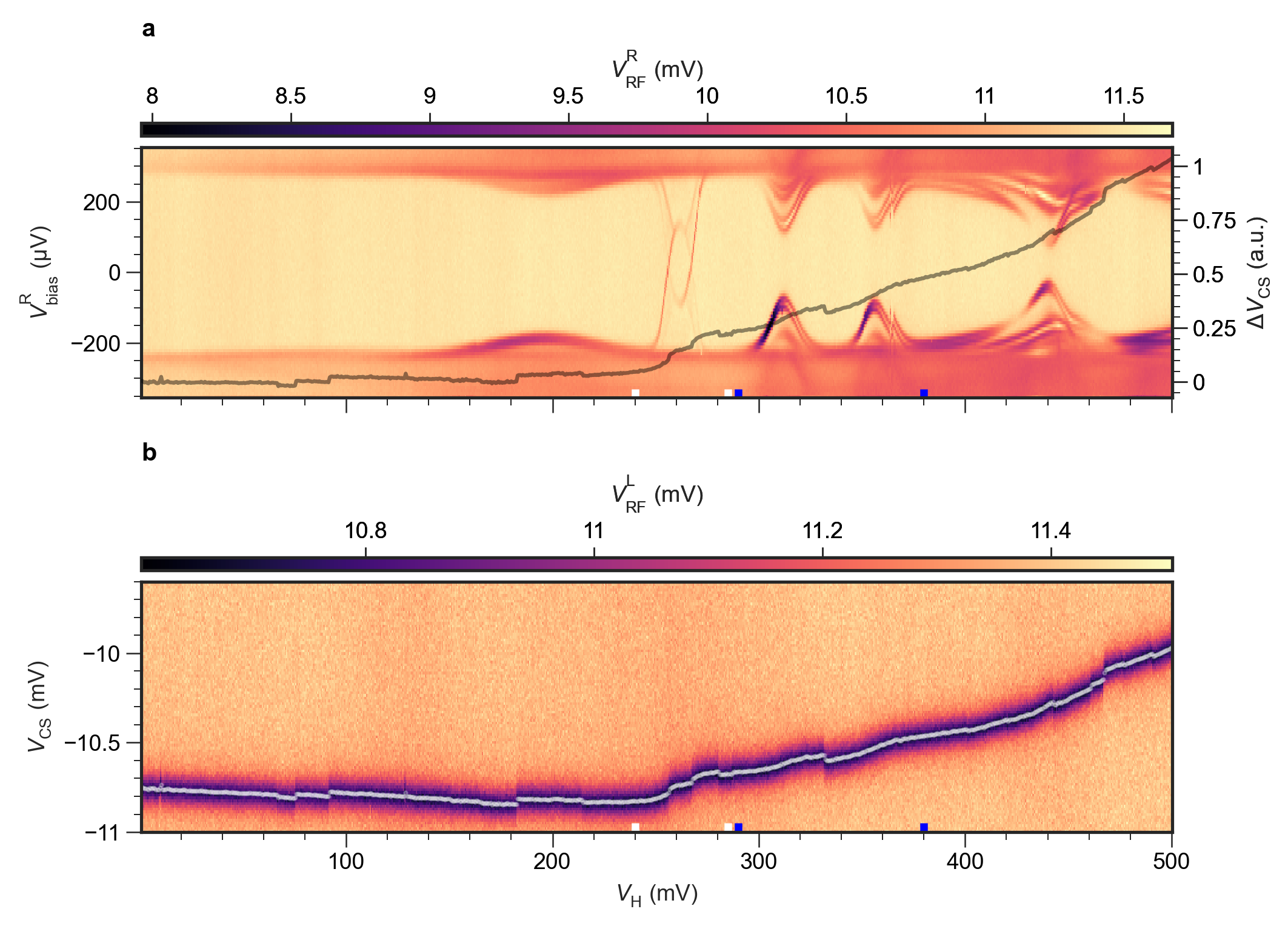}
    \caption{\textbf{Extended $\VH$ range measurement of \cref{fig: 1}.} 
    \textbf{a.} Amplitude of the reflected RF signal of the right lead, $\VRFR$ for varying hybrid gate voltage $\VHv$ and right bias $\VR$ at $B = \SI{100}{mT}$, $\VTR=0$ and $\VRO = \SI{500}{mV}$. \textbf{Superimposed line.} Shift of the CS Coulomb resonance for each $\VHv$ setting, extracted from b. A global slope is subtracted for clarity.
    \textbf{b.} Amplitude of the reflected RF signal of the left lead, $\VRFL$, for varying $\VCS$ and $\VH$. White dots indicate the CS resonance found from peak-finding. The ABSs shown in \cref{fig: 1} are indicated by the blue marks. The ABS of \cref{fig: 2} is indicated by the white marks. Panel b shows the corresponding charge sensor measurement.
    }\label{S: long_range}
\end{figure*}

\cref{S: long_range}a shows spectroscopy of the hybrid for a larger range of $\VH$ than shown in \cref{fig: 1}c. The ABSs shown in \cref{fig: 1} are indicated by the blue marks. The ABS of \cref{fig: 2} is indicated by the white marks. Panel b shows the corresponding charge sensor measurement.

\subsection{\cref{fig: 2} Charge Sensor Data Processing} \label{sec:fig2 data processing} 

\begin{figure*}[ht!]
    \centering
    \includegraphics[width=\linewidth]{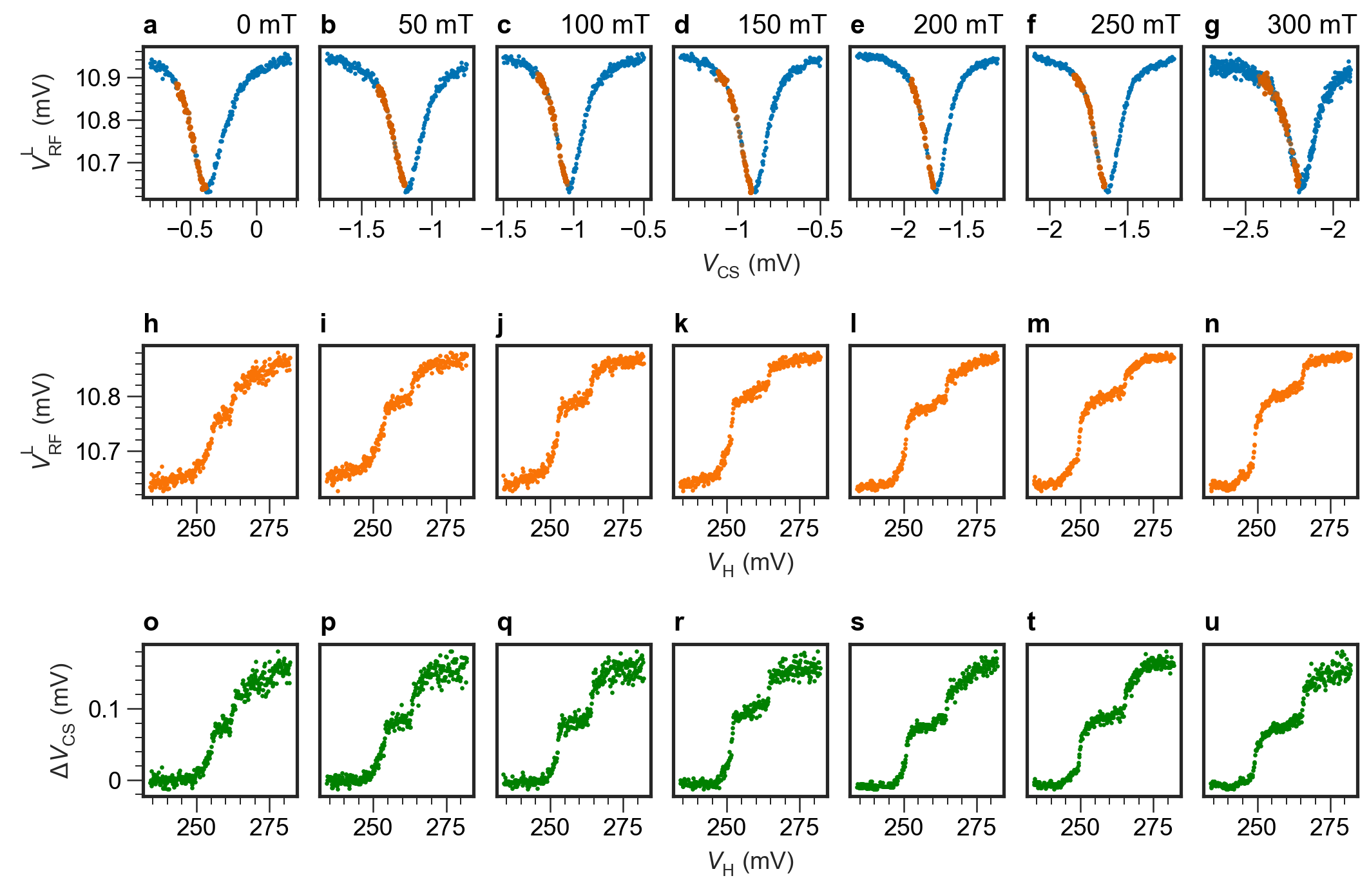}
    \caption{\textbf{Data processing of \cref{fig: 2}.} 
    \textbf{a-g.} Amplitude of the reflected RF signal of the left lead, $\VRFL$, for varying $\VCS$ at $\VH = \SI{234}{mV}$ for values of $B$ indicated in the title. These measurements are used to map a measured $\VRFL$ value to the closest $\VCS$ value. \textbf{Orange markers.} The $\VRFL$ values from the data shown in panels h-n, mapped to $\VCS$. For each $\VH$ setpoint, the measured $\VRFL$ is compared to the full $\VRFL$ v.s. $\VCS$ curve of panels a-g. Each measured $\VRFL$ value from panels h-n is then mapped to a $\VCS$ value from panels a-g.
    \textbf{h-n.} $\VRFL$ for varying $\VH$ at fixed $\VCS$ for values of $B$ indicated in the title. At each value of $\VH$, the measured $\VRFL$ is compared to the corresponding measurement of panels a-g. for the same $B$. The closest matching value of $\VCS$ is then chosen, resulting in a mapping of $\VH$ to $\VCS$, which is plotted in orange in panels a-g.    
    \textbf{o-u.} Shift of the gate voltage corresponding to the Coulomb resonance of the CS, $\dVCS$, for varying $\VH$. These correspond to the orange markers of panels a-g. after subtracting a global slope.
    }\label{S: data_process_fig2}
\end{figure*}

A change of the ABS charge results in a linear shift of the CS electrochemical potential, proportional to their mutual capacitance. The resulting change in reflected signal $\VRFL$ is not linear due to the shape of the Coulomb dip. To compensate for this, we first perform a characterization measurement of $\VRFL$ for varying $\VCSv$ for each $B$ value as shown in \cref{S: data_process_fig2}a-g. We then fix the $\VCSv$ gate value such that the CS is on Coulomb resonance and measure $\VRFL$ while varying $\VHv$, which is shown in \cref{S: data_process_fig2}h-n. For each measured $\VRFL$ value per $\VH$, we find the closest $\VRFL$ value in the characterization measurement. We then map the measured $\VRFL$ value to $\VCS$, which is indicated by the orange markers in \cref{S: data_process_fig2}a-g. Finally, we subtract a global slope from the resulting $\VCS$ value that we attribute to remaining cross-capacitance to $\VH$ and show the resulting $\Delta \VCS$ in \cref{S: data_process_fig2}o-u. We note that the vertical difference between the three plateaus is now roughly equal, as compared to \cref{S: data_process_fig2}h-n. Therefore, we interpret the processed data as being proportional to the ABS charge.
\FloatBarrier
\subsection{Andreev Bound States In the Atomic Limit} \label{sec:atomic limit} 
The inset of \cref{fig: 2}d shows the occupation of an ABS modeled in the atomic limit. In the many-body basis of a single orbital $\{\ket{0}, \ket{2}, \ket{\downarrow}, \ket{\uparrow} \}$ the Hamiltonian is:
$$H_\mathrm{ABS}= \begin{bmatrix}
0 & -\Gamma & 0 & 0 \\
-\Gamma & 2\mu+U & 0 & 0 \\
0 & 0 & \mu+E_\mathrm{Z} & 0 \\
0 & 0 & 0 & \mu - E_\mathrm{Z} \\
\end{bmatrix}  \label{supsec: atomic limit}$$
Where $\mu$ is the electrochemical potential of the uncoupled orbital, $\Gamma$ is the coupling to the superconductor, $U$ is the charging energy and $E_\mathrm{Z}$ is the Zeeman energy. We diagonalize the Hamiltonian, obtain the eigenvectors and calculate the ABS ground state occupation $\langle n \rangle$. For \cref{fig: 2}d, we use $\Gamma = 2.45$, $U=5$ and $E_\mathrm{Z} = 0, 0.25, 0.6$. For more details on the model, see Refs.~\cite{bauer_spectral_2007, martin-rodero_andreev_2012}.

For \cref{fig: 3}, we combine an atomic-limit ABS with a QD:
\begin{equation}
    H=H_\mathrm{ABS}+H_\mathrm{QD}+H_t
    \label{H_abs-qd}
\end{equation}
Where for the QD we have:
\begin{equation}
    H_\mathrm{QD} = (\mu_\mathrm{QD}-E_\mathrm{Z}) c_\downarrow^\dagger c_\downarrow + U_\mathrm{QD} c_{\uparrow}^\dagger c_{\uparrow} c_{\downarrow}^\dagger c_{\downarrow}.
\end{equation}
$H_t$ describes the tunnel coupling between the ABS and the QD and is given by:
\begin{equation}
    H_t = t(c_\downarrow^\dagger d_\downarrow + c_\uparrow^\dagger d_\uparrow + h.c.)+t_\mathrm{so}(-c_\downarrow^\dagger d_\uparrow + c_\uparrow^\dagger d_\downarrow + h.c.),
    \label{H_tunnel_ABS_QD_system}
\end{equation}
Where $t$ is the spin-conserving tunnelling amplitude, and $t_\mathrm{so}$ is the spin-flipping tunneling amplitude resulting from spin-orbit interaction.

The basis of \cref{H_abs-qd} can be split into even and odd parity subspaces, where we consider the following states in the $\ket{N, M}=\ket{N}_\mathrm{QD} \otimes \ket{M}_\mathrm{ABS}$ basis:  

\begin{align*}
    & \{\ket{0,S},\ket{2,S},\ket{\downarrow,\downarrow}\} &\mathrm{even}\\
    &\{\ket{0,\downarrow},\ket{\downarrow,S},\ket{2,\downarrow}\} &\mathrm{odd}
\end{align*}
Here $\ket{S} =  u\ket{0}-v\ket{2}$ is the ABS singlet. Using \cref{H_tunnel_ABS_QD_system}, we can calculate the effective coupling terms between the basis states of the system. We only consider transitions with a fixed global parity, i.e. only transitions between two odd occupation states or transitions between two even occupation states.
The effective coupling between the different even occupation states of the system is given by:
\begin{equation}
    \braket{2,S|H_t|0,S}=0
    \label{1}
\end{equation}
\begin{equation}
    \braket{\downarrow,\downarrow|H_t|0,S}=vt_\mathrm{so}
    \label{2}
\end{equation}
\begin{equation}
    \braket{2,S|H_t|\downarrow,\downarrow}=-ut_\mathrm{so}
    \label{3}
\end{equation}
For the odd occupation states we have:
\begin{equation}
    \braket{0,\downarrow|H_t|\downarrow,S}=ut
    \label{4}
\end{equation}
\begin{equation}
    \braket{0,\downarrow|H_t|2,\downarrow}=0
    \label{5}
\end{equation}
\begin{equation}
    \braket{2,\downarrow|H_t|\downarrow,S}=vt
    \label{6}
\end{equation}
We can now write down \Cref{H_abs-qd} in the even and odd parity subspace matrix representation:
\begin{equation*}
    H_{even}=\begin{bmatrix}
        E_\mathrm{S} & 0 & v t_\mathrm{so} \\
        0 &  2\mu_\mathrm{QD}+U_\mathrm{QD}+E_\mathrm{S}& -ut_\mathrm{so} \\
        vt_\mathrm{so} & -ut_\mathrm{so} & \mu_\mathrm{QD}-E_\mathrm{Z} + E_\downarrow\\    
    \end{bmatrix}
\end{equation*}and 
\begin{equation*}
    H_{odd}=\begin{bmatrix}
        E_\downarrow&ut&0\\
        ut&\mu_\mathrm{QD}-E_\mathrm{Z}+E_\mathrm{S}&vt\\
        0&vt&2\mu_\mathrm{QD}+U_\mathrm{QD}+E_\downarrow\\ 
    \end{bmatrix}
\end{equation*}
Here $E_\mathrm{S}$ and $E_\downarrow$ are the ABS singlet and doublet energies respectively. The full many-body matrix describing the system is given by the following block-diagonal matrix:

\begin{equation}
    H_{total}=\begin{bmatrix}
        H_{even} & 0 \\
        0 & H_{odd} \\
    \end{bmatrix}
    \label{total_matrix}
\end{equation}
The ground state of the system corresponds to the eigenstate of the lowest eigenenergy of \cref{total_matrix}. \cref{fig: 3}b was computed for $U_\mathrm{ABS} = 0.03$, $U_\mathrm{QD} = 1$, $\Gamma = 0.3$, $E_\mathrm{Z} = 0.7$, $t=0.15$, $t_\mathrm{SO}=0.01$.

\subsection{\cref{fig: 3} DC Transport and RF reflectometry comparison}\label{sec:DC fig 3} 
\begin{figure*}[ht!]
    \centering
    \includegraphics[width=\linewidth]{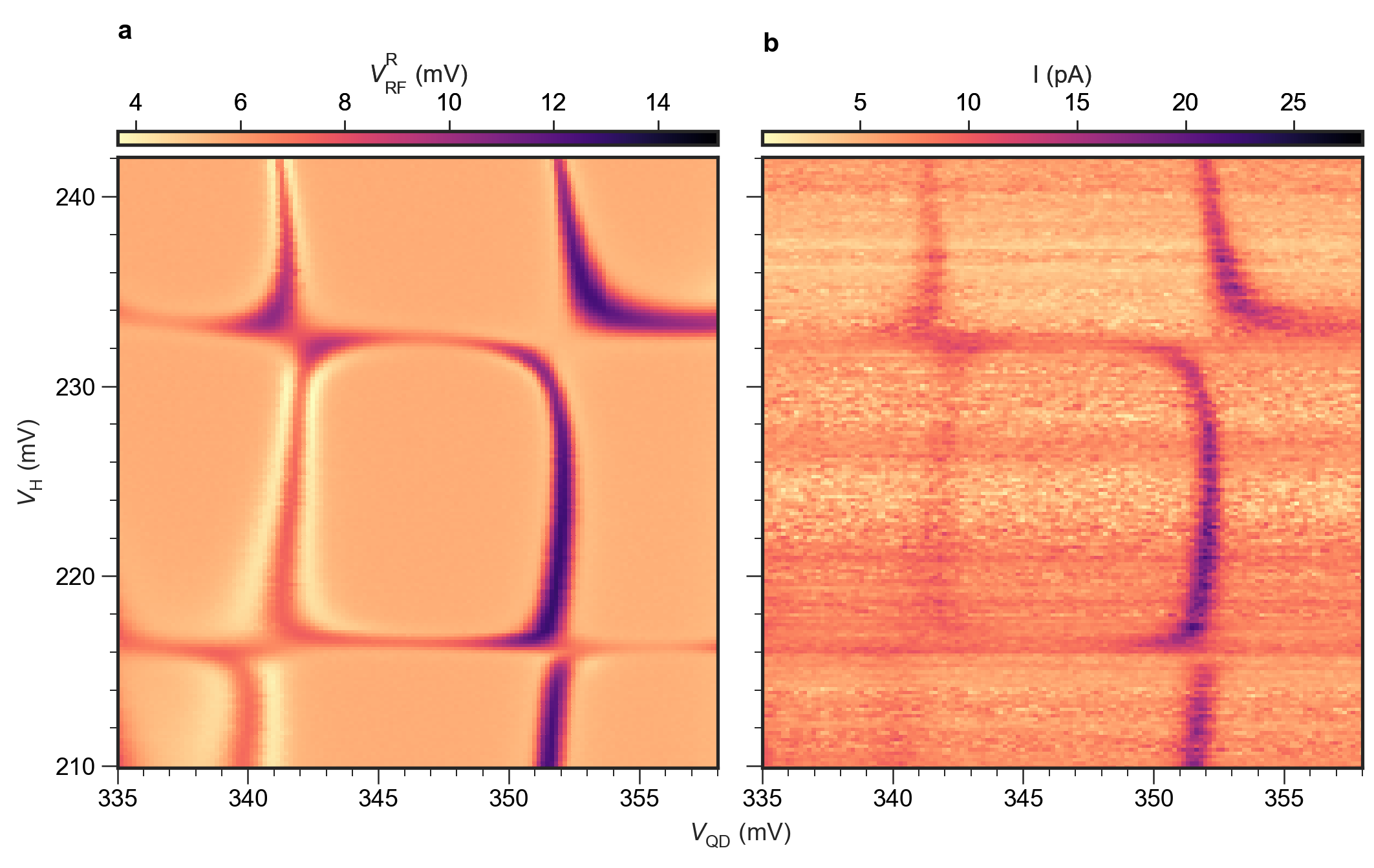}
    \caption{\textbf{Comparison of DC transport and RF reflectrometry of \cref{fig: 3}. at $\VR=0$.} 
    \textbf{a.} Amplitude of the reflected RF signal of the right lead, $\VRFR$ for varying QD plunger gate voltage, $\VQD$, and $\VH$. 
    \textbf{a.} Current measured at the right lead, $I$, for varying QD plunger gate voltage, $\VQD$, and $\VH$. 
    }\label{S: RF_DC_fig3}
\end{figure*}

In \cref{S: RF_DC_fig3} we compare the charge stability diagram of \cref{fig: 3}a as measured in RF reflectometry and DC transport. Most of the features seen in \cref{S: RF_DC_fig3}a are also seen in \cref{S: RF_DC_fig3}b. In $\VRFR$, some features are more visible than in DC transport. While there is only a finite current when the QD undergoes Andreev reflection or  hybridizes with the ABS, there is always a dip in $\VRFR$ when the normal lead and QD are on resonance. This can be explained by the excess dissipation known as "Sisyphus resistance", which originates from the QD level and Fermi level of the normal lead being detuned with an AC voltage~\cite{persson_excess_2010}. We attribute the measured current at zero-bias in \cref{S: RF_DC_fig3}b to a finite voltage offset.

\subsection{Charge Sensor SNR}\label{sec:SNR} 
\begin{figure*}[ht!]
    \centering
    \includegraphics[width=\linewidth]{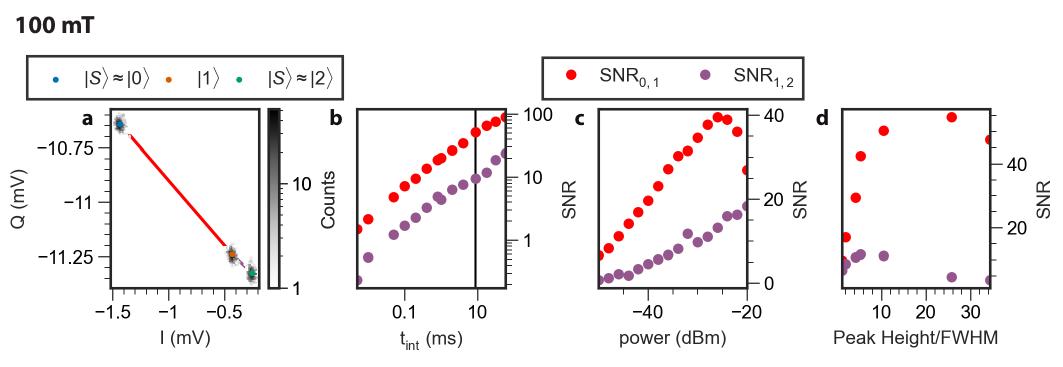}
    \caption{\textbf{Charge Sensor SNR measurements for a single ABS.} 
    \textbf{a.} Example histograms of the in-phase response and quadrature, I and Q of the left resonator, for the 0-like even, 1 and 2-like even occupations of the ABS. The colored markers indicate the charge state of the ABS that corresponds to each histogram. The red and purple lines show the distance between the histograms that we use to calculate the SNR.
    \textbf{b.} The SNR calculated from IQ-histograms for the 0-like singlet and doublet states (red), and 2-like singlet and doublet states (purple) for varying integration time $t_\mathrm{int}$. The vertical black line indicates, $t_\mathrm{int} = \SI{9.3}{ms}$, which is the integration time used in the main text. The RF out signal had a power of \SI{-30}{~dBm} for every $t_\mathrm{int}$.
    \textbf{c.} Same as panel b., for varying power of the RF input signal and at fixed $t_\mathrm{int} = \SI{9.3}{ms}$.
    \textbf{d.} Same as panel c., for varying shape of the CS Coulomb resonance at fixed $t_\mathrm{int} = \SI{9.3}{\milli\second}$ and power \SI{-30}{~dBm}. Each datapoint is taken for a different gate voltage $\VLO$, resulting in Coulomb peaks of different width. We present the SNR versus Coulomb peak height, divided by the full width at half maximum.
    }\label{S: SNR}
\end{figure*}

To estimate the SNR of the charge sensor, we measure the in-phase response and quadrature, I and Q, of the left resonator for the ABS of \cref{fig: 2} at $B=\SI{100}{mT}$. We measure the I and Q signals over time at different $\VH$ values and plot the result in histograms, as shown in \cref{S: SNR}a. For each histogram, we compute the average IQ response $\mu_{i}$ and the standard deviation $\sigma_i$. Then, we define the SNR for two charge states as: $$\mathrm{SNR}_\mathrm{i, j} =\frac{|\mu_i-\mu_j|}{\sigma_i + \sigma_j}$$ 
See Supplemental Material of Ref.~\cite{de_jong_rapid_2021} for details. The red and purple lines in \cref{S: SNR}a show $|\mu_i-\mu_j|$ for the 0 and 1, and 1 and 2 occupations of the ABS. Dividing by the sum of the standard deviations results in the SNR. 
\cref{S: SNR}b shows SNR calculated for varying integration time, $t_\mathrm{int}$. \cref{S: SNR}c shows SNR calculated for varying RF output power. It reaches a maximum of $39.5$ for -26 dBm for the SNR of the 0-like singlet and 1 occupations. \cref{S: SNR}d shows SNR calculated for varying tunnel gate voltage $\VLO$. We fit the resulting Coulomb peak and extract the ratio of peak height and full width at half maximum. We see that a sharper Coulomb peak results in a higher SNR.

\end{document}